This article is under review.
Content starts from the next page.

# A Survey of User Expectations
# and Tool Limitations
# in Collaborative Scientific Authoring and
# Reviewing

Afshin Sadeghi[1], Mahdi Jaberzadeh Ansari[1], Johannes Wilm[3], and Christoph
Lange[1,2]

[1] Institute for Applied Computer Science, University of Bonn, Germany
[2] Fraunhofer Institute f. Intelligent Analysis and Information Systems IAIS, Germany
[3] GESIS – Leibniz Institute for the Social Sciences, Germany
{sadeghi,langec}@cs.uni-bonn.de,{mahdi.jbz,johanneswilm}@gmail.com

**Abstract.** Collaborative scientific authoring is increasingly being sup-
ported by software tools. Traditionally, desktop-based authoring tools
had the most advanced editing features, allowed for more formatting
options, and included more import/export filters. Web-based tools have
excelled in their collaboration support. Recently, developers on both sides
have been trying to close this gap by extending desktop-based tools to
better support collaboration and by making web-based tools richer in
functionality. To verify to what extent these developments actually meet
the needs of researchers, we gathered precise requirements towards better
tool support for scientific authoring and reviewing workflows by inter-
viewing 213 users and studying a corpus of 27 documents. We present
the design of the survey and interpret its results. The conclusion is that
WYSIWYG and offline desktop authoring tools continue to be more pop-
ular among academics than text-based and online editors.

## 1 Introduction

In the OSCOSS research project on Opening Scholarly Communication in Social
Sciences[4], we have developed an integrated framework [5] to support authoring
and direct review and discussion by providing chat and commenting facilities in
the environment in which a paper is authored. To identify the most important
features and requirements toward our goal, we conducted a study asking scholars
about their experience with the current authoring systems and identifying the
actual usage of certain document features in a corpus of scientific papers.

---

[4] https://github.com/OSCOSS, http://eis.iai.uni-bonn.de/Projects/OSCOSS

For the purpose of this study, it made sense to distinguish tools installed on the user's computer as ordinary software (desktop tools) from those that run in the browser (web-based tools). While these terms may not be the most descriptive, they are nevertheless widely used within the industry.

We present an overview of related work in section 2. We then describe the survey method in section 3. Next, we present the results of the user survey in section 4. An evaluation of a corpus of papers and their features is presented in section 5. We conclude with an outlook in section 6.

## 2 Related Work

Related work includes research on aspects of authoring tools that are relevant for academics, as well as conducting online surveys using questionnaires.

Thoma et al. evaluate **authoring** in their interactive publication system [7], measured the ease of creation and whether specialized skills are required to use the tool and the skill level necessary. In a historical review and analysis of authoring tools [4], Locatis and Al-Nuaim describe that an 'easier' authoring tool not only increases output and users but also improves quality and enhances productivity. In analyzing authoring requirements, they argue that the choice of an authoring tool should not be grounded in comparing the number or quality of software features because it tends to give preference to feature-heavy tools rather than the tool with the *relevant* features. In their evaluation they focus on the support of desired outputs, different platforms, proper layouts, and multimedia aspects, as well as ease of use and cost.

For the **survey**, we were interested in an Internet questionnaire due to the ease with which it would permit us to do a multi-country study. Wright [8] explains that online surveys have the advantage of addressing unique populations. He specially argues that because of the popularity of the Internet, online surveys have permitted covering groups and individuals who previously were difficult to reach. The online medium of a survey saves time and cost because it allows a researcher to reach a relatively large number of people with common characteristics within a short amount of time, and the cost of online surveys is much lower than that of classic paper-and-pencil surveys. However, online surveys have disadvantages and issues. For example, the tone used in online communication in such surveys could be considered rude or offensive by certain communities [3], or the emails sent out could up being classified as spam [1]. Also, some participants in online surveys may be more biased than what will normally be found in traditional surveys, due to the inherent limitation of options to pick a representative sample [6]. Also, any online survey cuts off a part of the population that is not accessible online – in our case, possibly, some users preferring desktop-based tools.

## 3 Survey Methodology

This section describes the different stages of our survey methodology. To create a sample population of individuals with authoring experience we first extracted 6,000 email addresses of scholars from the websites of 60 different universities worldwide. We emailed these people and picked those who consented to participate in the survey and had authoring experience as our sample population.

Our method of data collection was a two step procedure. In the first step, we created a questionnaire that focused on what the key requirements of authoring tools were for the authors. We conveyed our designed questions to the survey participants by means of an online questionnaire platform [5]. In the second step, we compared different authoring tools to find out how well they covered the key requirements of academics we had gathered in the first step.

### 3.1 The Sample Population

Totally 213 persons participated in our online survey. They were living in 15 different countries and came from 60 different universities. 92% of the participants were graduate students or in higher education. The level of education of our varies across the population sample. More than half of them were from the field of Information Technology and Mathematics, and Professors with about 32% of sample individuals form the biggest group of participants.

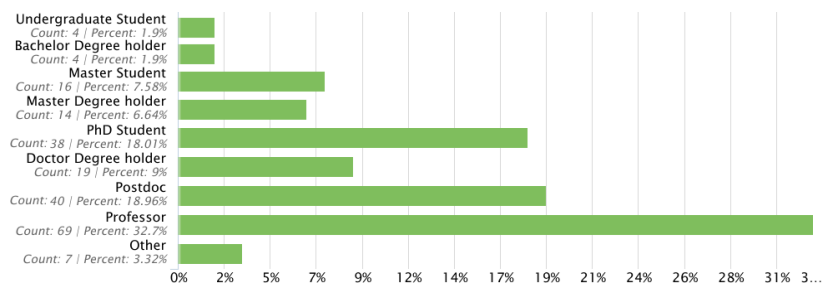

Fig. 1: Educational level of the participants in our online survey.

Those participants with the experience of collaborative scientific writing and peer reviewing were most relevant to this survey; therefore, we asked the interviewees whether they had ever participated in a collaboration to produce a piece of scientific writing, and whether they had ever been involved in a peer review.

As illustrated in Figure 4a and Figure 4b, more than 80% of the participants had experienced collaborative writing and more than 78% of them have performed peer reviewing at least once. Knowing the percentages of the experts in these two areas is crucial because it demonstrates that we had interviewed the right sample of the population.

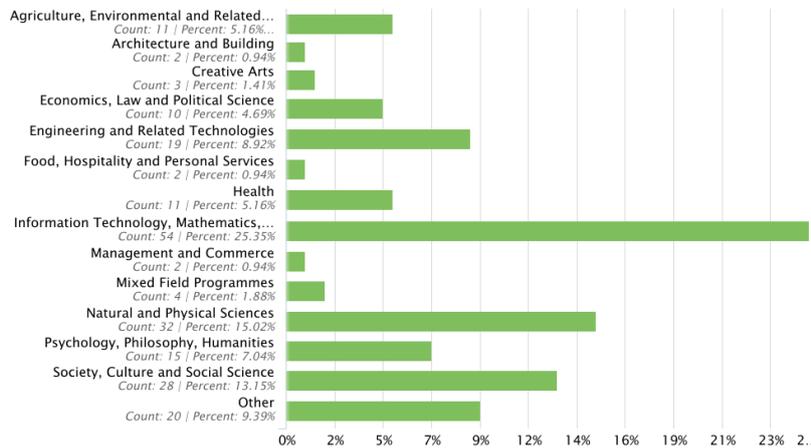

Fig. 2: Distribution of the field of study of the participants in our online survey.

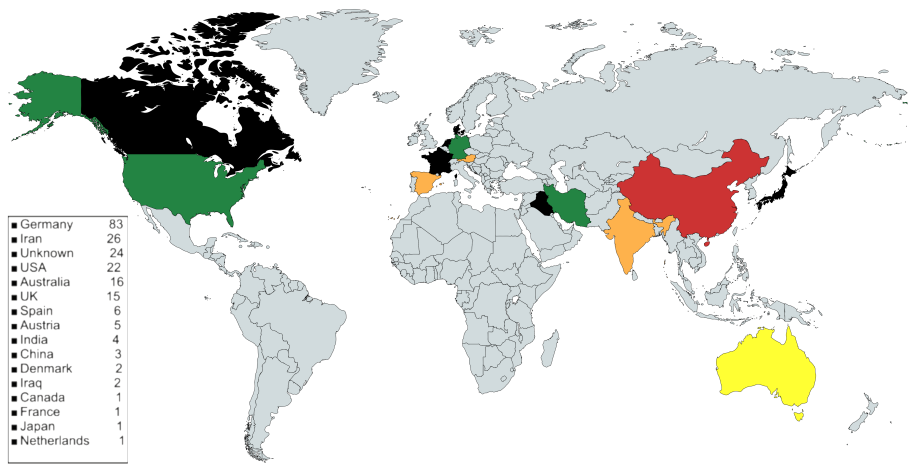

Fig. 3: Distribution of the participants in our online survey in a world map.

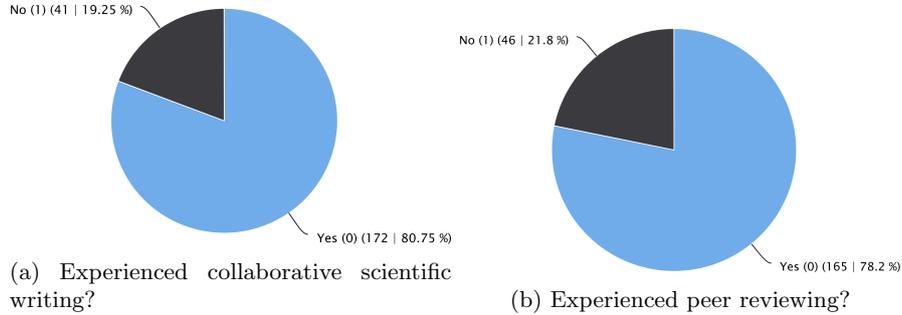

(a) Experienced collaborative scientific writing?

(b) Experienced peer reviewing?

Fig. 4: Experience of the participants

## 4 Tools Selection Survey

We directed our survey toward finding those tools that the participants found most useful and usable. We listed the tools to be considered to survey and the tools that the participants used and evaluated the functionality of the tools from their point of view. We later inferred their experience and expectations from this functionality.

| Category | Type | Votes |
| --- | --- | --- |
| Microsoft Word | Desktop | 182 |
| Google Docs | Web | 106 |
| TeX/LaTeX | Desktop | 93 |
| Microsoft Office 365 | Web | 42 |
| Microsoft Office Word Online | Web | 34 |
| LibreOffice Writer | Desktop | 29 |
| Overleaf | Web | 20 |
| ShareLaTeX | Web | 18 |
| Apache OpenOffice Writer | Desktop | 15 |
| Mathematica | Desktop | 15 |
| WordPerfect | Desktop | 14 |
| EtherPad | Web | 8 |
| LyX | Desktop | 8 |
| Other from the list | Desktop | 23 |
| Other from the list | Web | 13 |
| Other unknown (Desktop & Web-based) | Desktop, Web | 52 |

Table 1: Number of votes for different authoring tools.

To find out which authoring software was most popular among scientists for scientific writing, we first created a list of tools that we estimated to be well-known and that were also mentioned as TeX editors or word processors on

Wikipedia at the time of the survey[6]. We then asked the interviewees which of these authoring tools they had used for scientific writing in the past. They were allowed to select more than one tool. Afterward, we removed those tools from the list that none of our interviewees had mentioned (see Table 1). The responses showed a clear preference for desktop over web-based tools.

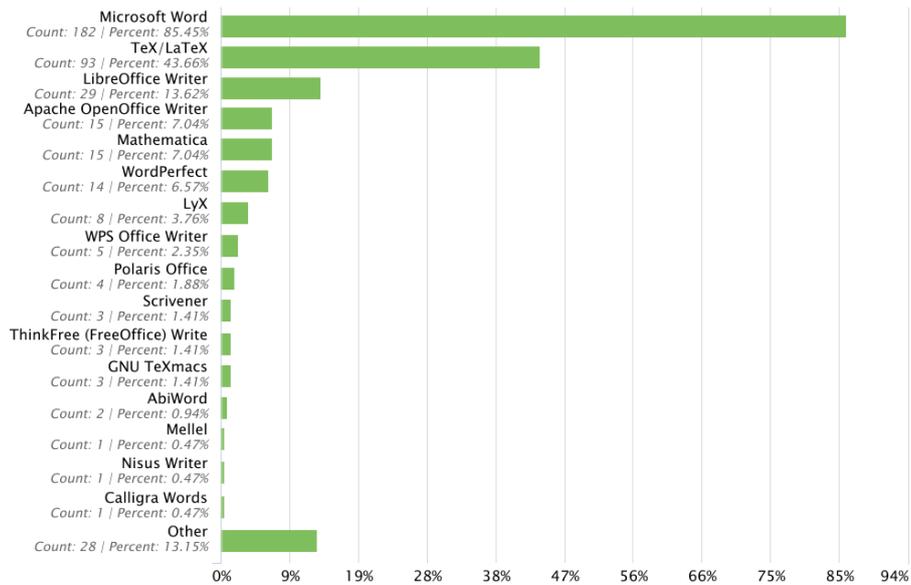

Fig. 5: Number of interviewees using desktop authoring tools for scientific writing

We asked the participants which of the tools they had used in the past. This time we presented them with separate lists for desktop (Figure 5) and web-based (Figure 6) tools. The results showed similar tendencies as the answers to the previous question (Table 1), as well as that the most popular authoring tool was Microsoft Word among the desktop software and Google Docs among the web-based authoring tools.

We further asked the interviewees to rate each of the tools they used on a scale 0 to 5 according to two criteria: Overall quality of the tool and whether they were satisfied with the tool fulfilling their own requirements. Figure 7 displays the questionnaire results sorted based on the number of the total votes. It shows that, on average, the sample population was more satisfied with Microsoft Word than Google Docs. Furthermore, even between the desktop versions of Microsoft Word and web-based ones, the participants preferred desktop ver-

---

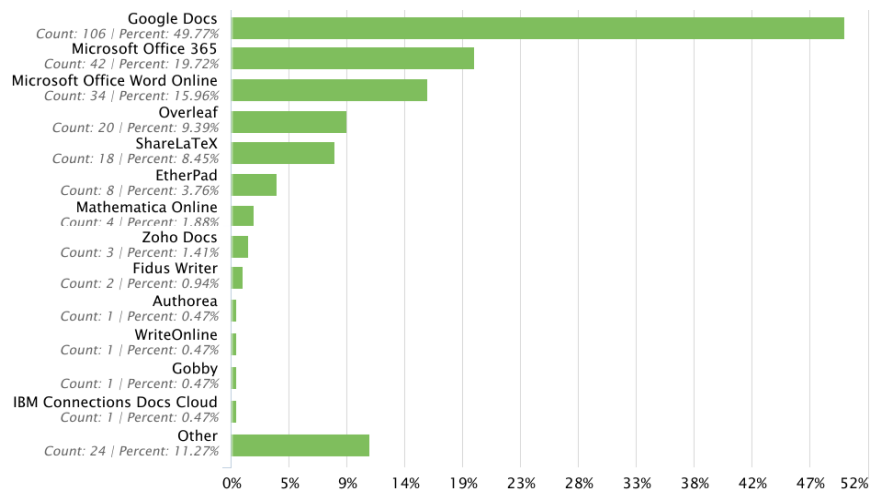

Fig. 6: Number of interviewees users using web-based authoring tools for scientific purposes

sions. Also, WYSIWYG authoring tools were apparently more popular than text-based tools like LaTeX editors. Even though Authorea tried to market itself as a "Google Docs for Academics"[7], there was not a significant percentage of our interviewed scientists who had used Authorea. The fact that Authorea was a WYSIWYG editor did not help it in gaining popularity among the sample population. A problem with Authorea at the time of the survey in Nov 2016 was that its WYSIWYG interface required a user to press an extra key to start a new paragraph. The user was required to use the 'Insert below' or 'Edit' buttons to be able to add a new paragraph or edit an existing one. These extra clicks reduced the ease of use of the product despite it having an effective citation insertion mechanism. These steps seem to be eliminated in the recent versions of the Authorea app.

## 5 Features

To find out which features were most relevant to social science authors, we analyzed a corpus of 27 social science articles in DOCX format and extracted the features that were most used in these articles. Afterwards, we asked the interviewees about their favorite features and limitations in general and compared their perceived limitations and the actual features extracted from papers. We finally evaluated to what extent these tools support the most favored features.

[7] https://techcrunch.com/2014/09/22/authorea-seed/

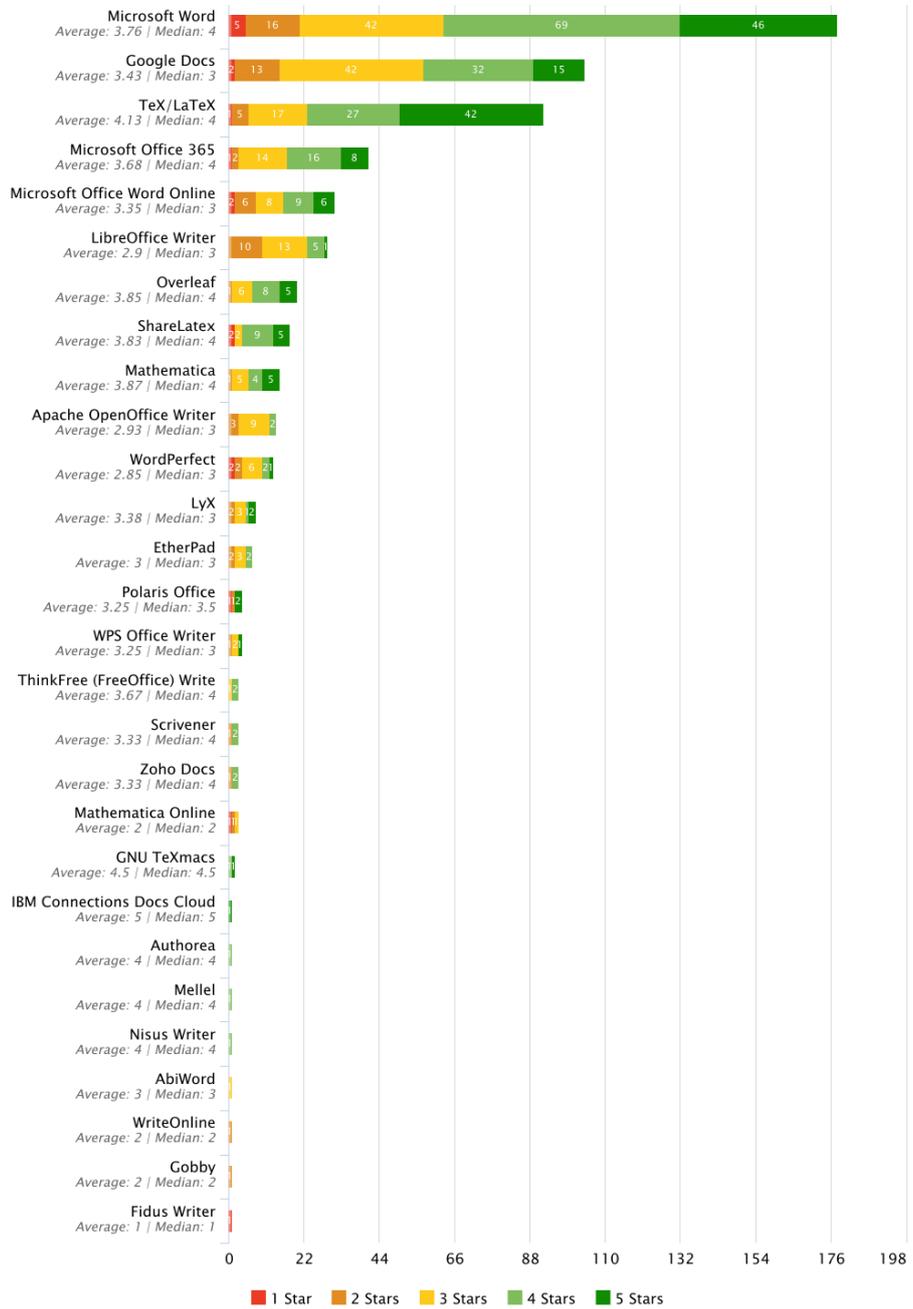

Fig. 7: Results of the satisfaction ratings on the different authoring tools

### 5.1 Feature Extraction

We looked through the 27 scientific articles in DOCX format in to find out what standard features[8] were most frequently used and if their usage had followed the intended purpose. We observed that abbreviations and citation management were the major feature that was most widely used [9] in our evaluated sample documents. Other extracted features are available to ref [2]. A good example of an efficient citation support was the citation facility of Authorea. Authorea allowed the typing of a part of the name of an author, a keyword or the year of publication. It would then provide a list of possible items, and with one click these could be imported into the document. This facility seemed helpful for authors who do not have to type each property of a source into a form.

There are some citation add-ons for Google Docs and Microsoft Word, however, as we tested them, they were not as efficient and in most of the cases could not find any related item. In Microsoft Word, there was a built-in citation manager, but it was limited and required at least 5 fields to be entered manually, which is time-consuming. We observed during our investigation of the real scientific papers that in most cases people preferred to insert citations as simple text from other citation tools like Google Scholar or Mendeley. Importing the citation as a simple text from a trustworthy source was favored, likely because there were fewer typos in the citations when entering them using a tool rather than writing citations completely manually. Manually maintaining such citation list is difficult as it requires several editing steps when a citation that needs to be updated occurs in different parts of a document.

### 5.2 User Evaluation

The most significant observation in 5.1 is in line with a majority of interviewees who also mentioned citation as the most important feature in scientific writings (Figure 8).

The wording of the question to the interviewees about the most important feature was as follows:

– In your opinion, what are the most important features for an authoring tool to have in order to be effective for use in the collaborative writing of scientific papers?

One can observe in Figure 8 that almost 50% of the interviewees expected the authoring tool to support exporting to and importing from other formats. Figure 8 also indicates that supporting real-time collaboration was preferred over offline collaboration. Some of the interviewees mentioned the ability to track changes as their favorite feature. When a tool provides real-time collaboration tracking the changes of the document becomes useful. Google Docs supports change tracking by displaying the changes of the document from a window of a few seconds. The observed data indicated that providing an efficient tool for citation with the ability of online search in bibliography databases probably increases the popularity of an authoring tool in scientific communities, although, as the example of Authorea shows, it does not guarantee that such tool be chosen over others. We queried the interviewees about their experience with reviewing and their expectations from these tools. We asked them which of the methods or software tools they would prefer for peer reviewing. We then asked them which of the features were most important for a specialized reviewing software tool.

---

[8] Classification of DOCX Standard Scientific Style, `http://docx.science.ai`

[9] Abbreviations and the citations respectively appeared 3,026 and 1,232 times in the assessed DOCX files.

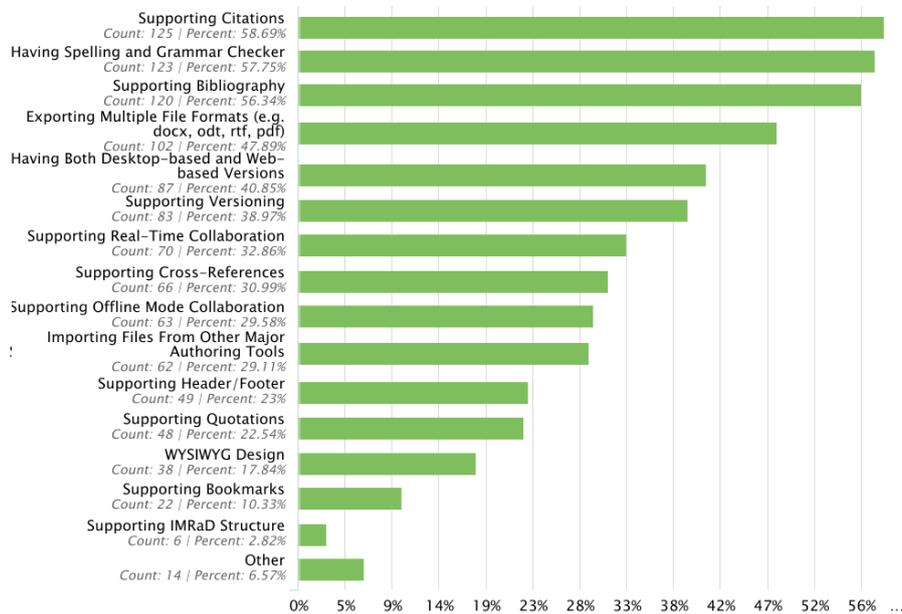

Fig. 8: Opinion of the interviewees about the most important features of an authoring tool, which is suitable for scientific writings.

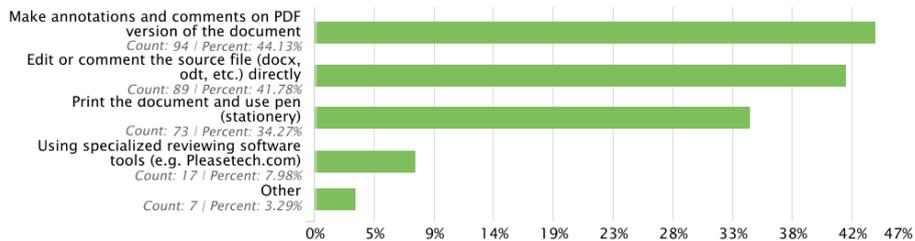

Fig. 9: Usual methods that the interviewees used for the peer reviewing

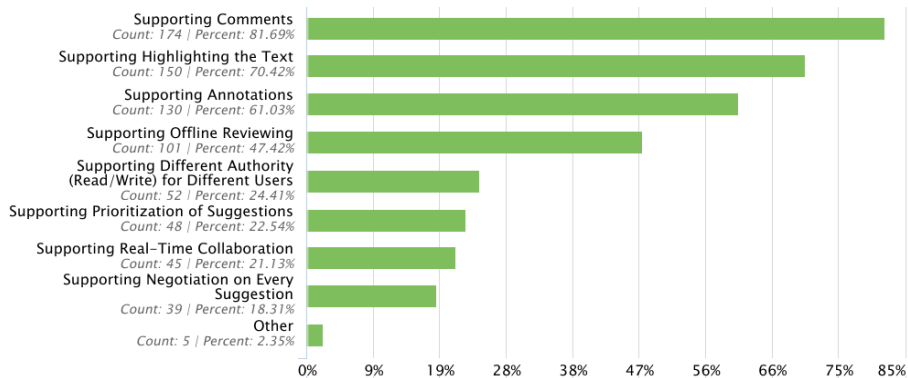

Fig. 10: Opinions of the interviewees about the most important features for a specialized reviewing software tool

Figures 9 and 10 illustrate the result of the above questions. It can be seen that PDF tools play a considerable role in reviewing documents – for example, Adobe Reader or Foxit Reader thanks to their annotation and commenting features. In Figure 9 we also observe that almost 42% of respondents knew that some tools like Microsoft Word have reviewing options. Nevertheless, 34% preferred to print the documents and use a pen for reviewing. Using a print+pen solution means that much more time will be needed to digitize suggestions and the communication between authors and reviewers. Furthermore, choosing this method means we cannot track changes and perceive the evolution of the document.

Finally, it can be seen that just 8% were interested in using extra tools for reviewing. This indicates that social science academics would tend to prefer using an integrated tool that supports authoring and reviewing together rather than two individual tools.

Among the expectations of our interviewed sample population from a reviewing tool, Figure 10 illustrates that support for commenting, highlighting and annotations were three major expectations among the interviewees. It can be seen that support for offline reviewing was more important than support for real-time reviewing, while real-time collaboration functionality was of more interest in authoring tools. Furthermore, almost 75% thought providing read/write access was not important and a less than 2% of the interviewees suggested change tracking as an important feature for a reviewing tool.

## 6 Conclusion

In this study, we surveyed academics from 60 different universities worldwide about their preference for editing tools for academic texts. We found out that the top features users requested, such as citation management, are not provided well by WYSIWYG word processors, but that users nevertheless prefer desktop word processors with more general features over script- or web-based specialized tools.

Our sample of academics may not be fully representative of all academics worldwide in all respects, but we have no reason to believe that there is significant bias when it comes to choice of editing tools.

We conducted this study as part of the Opening Scholarly Communication in the Social Sciences (OSCOSS) project in which we add features to the open source academic web-based editor Fidus Writer. The results of this survey has made us realize that beside the focus on features specific to academics, we need to ensure to support rich general text editing features to become an attractive alternative for social scientists.

## References

1. Andrews, D., Nonnecke, B., Preece, J. Electronic survey methodology: A case study in reaching hard-to-involve Internet users. In: International Journal of Human-Computer Interaction 16(2) (2003), 185–210.
2. Ansari, M. J. Supporting Collaborative Authoring and Reviewing Workflows in Desktop-based and Web-based Authoring Environments. An optional note. MA thesis. University of Bonn, 2016. `https://git.io/v9fl8`.
3. Hudson, J. M., Bruckman, A. "Go away:"? Participant objections to being studied and the ethics of chatroom research. The Information Society. In: The Information Society 20(2) (2004), 127–139.
4. Locatis, C., Al-Nuaim, H. Interactive technology and authoring tools: A historical review and analysis. In: Educational Technology Research and Development 47(3) (1999), pp. 63–75. `http://dx.doi.org/10.1007/BF02299634`.
5. Sadeghi, A. et al. Opening Scholarly Communication in Social Sciences by Connecting Collaborative Authoring to Peer Review. In: Information – Wissenschaft & Praxis (2017). arXiv: `1703.04428 [cs.DL]`. In press.
6. Stanton, J. M. An Empirical Assessment of Data Collection using the Internet. In: Personnel Psychology 51(3) (1998), pp. 709–725.
7. Thoma, G. R. et al. Interactive Publication: The document as a research tool. In: Web Semantics (Online) 8(2-3) (2010), 145–150.
8. Wright, K. B. Researching Internet-Based Populations: Advantages and Disadvantages of Online Survey Research, Online Questionnaire Authoring Software Packages, and Web Survey Services. In: Journal of Computer-Mediated Communication 10(3) (2005), pp. 00–00. `http://dx.doi.org/10.1111/j.1083-6101.2005.tb00259.x`.